%% file: paper.tex
\documentclass[letterpaper,compsoc,twoside]{IEEEtran}
\usepackage{fixltx2e} 
\usepackage{cmap} 
\usepackage{ifthen}
\usepackage[T1]{fontenc}
\usepackage[utf8]{inputenc}
\usepackage{amsmath}

\usepackage[font={small,it},labelfont=bf]{caption}
\usepackage{float}

\pdfoutput=1
\usepackage{scipy}
\makeatletter
\def\PY@reset{\let\PY@it=\relax \let\PY@bf=\relax%
    \let\PY@ul=\relax \let\PY@tc=\relax%
    \let\PY@bc=\relax \let\PY@ff=\relax}
\def\PY@tok#1{\csname PY@tok@#1\endcsname}
\def\PY@toks#1+{\ifx\relax#1\empty\else%
    \PY@tok{#1}\expandafter\PY@toks\fi}
\def\PY@do#1{\PY@bc{\PY@tc{\PY@ul{%
    \PY@it{\PY@bf{\PY@ff{#1}}}}}}}
\def\PY#1#2{\PY@reset\PY@toks#1+\relax+\PY@do{#2}}

\expandafter\def\csname PY@tok@gd\endcsname{\def\PY@tc##1{\textcolor[rgb]{0.63,0.00,0.00}{##1}}}
\expandafter\def\csname PY@tok@gu\endcsname{\let\PY@bf=\textbf\def\PY@tc##1{\textcolor[rgb]{0.50,0.00,0.50}{##1}}}
\expandafter\def\csname PY@tok@gt\endcsname{\def\PY@tc##1{\textcolor[rgb]{0.00,0.27,0.87}{##1}}}
\expandafter\def\csname PY@tok@gs\endcsname{\let\PY@bf=\textbf}
\expandafter\def\csname PY@tok@gr\endcsname{\def\PY@tc##1{\textcolor[rgb]{1.00,0.00,0.00}{##1}}}
\expandafter\def\csname PY@tok@cm\endcsname{\let\PY@it=\textit\def\PY@tc##1{\textcolor[rgb]{0.25,0.50,0.56}{##1}}}
\expandafter\def\csname PY@tok@vg\endcsname{\def\PY@tc##1{\textcolor[rgb]{0.73,0.38,0.84}{##1}}}
\expandafter\def\csname PY@tok@vi\endcsname{\def\PY@tc##1{\textcolor[rgb]{0.73,0.38,0.84}{##1}}}
\expandafter\def\csname PY@tok@mh\endcsname{\def\PY@tc##1{\textcolor[rgb]{0.13,0.50,0.31}{##1}}}
\expandafter\def\csname PY@tok@cs\endcsname{\def\PY@tc##1{\textcolor[rgb]{0.25,0.50,0.56}{##1}}\def\PY@bc##1{\setlength{\fboxsep}{0pt}\colorbox[rgb]{1.00,0.94,0.94}{\strut ##1}}}
\expandafter\def\csname PY@tok@ge\endcsname{\let\PY@it=\textit}
\expandafter\def\csname PY@tok@vc\endcsname{\def\PY@tc##1{\textcolor[rgb]{0.73,0.38,0.84}{##1}}}
\expandafter\def\csname PY@tok@il\endcsname{\def\PY@tc##1{\textcolor[rgb]{0.13,0.50,0.31}{##1}}}
\expandafter\def\csname PY@tok@go\endcsname{\def\PY@tc##1{\textcolor[rgb]{0.20,0.20,0.20}{##1}}}
\expandafter\def\csname PY@tok@cp\endcsname{\def\PY@tc##1{\textcolor[rgb]{0.00,0.44,0.13}{##1}}}
\expandafter\def\csname PY@tok@gi\endcsname{\def\PY@tc##1{\textcolor[rgb]{0.00,0.63,0.00}{##1}}}
\expandafter\def\csname PY@tok@gh\endcsname{\let\PY@bf=\textbf\def\PY@tc##1{\textcolor[rgb]{0.00,0.00,0.50}{##1}}}
\expandafter\def\csname PY@tok@ni\endcsname{\let\PY@bf=\textbf\def\PY@tc##1{\textcolor[rgb]{0.84,0.33,0.22}{##1}}}
\expandafter\def\csname PY@tok@nl\endcsname{\let\PY@bf=\textbf\def\PY@tc##1{\textcolor[rgb]{0.00,0.13,0.44}{##1}}}
\expandafter\def\csname PY@tok@nn\endcsname{\let\PY@bf=\textbf\def\PY@tc##1{\textcolor[rgb]{0.05,0.52,0.71}{##1}}}
\expandafter\def\csname PY@tok@no\endcsname{\def\PY@tc##1{\textcolor[rgb]{0.38,0.68,0.84}{##1}}}
\expandafter\def\csname PY@tok@na\endcsname{\def\PY@tc##1{\textcolor[rgb]{0.25,0.44,0.63}{##1}}}
\expandafter\def\csname PY@tok@nb\endcsname{\def\PY@tc##1{\textcolor[rgb]{0.00,0.44,0.13}{##1}}}
\expandafter\def\csname PY@tok@nc\endcsname{\let\PY@bf=\textbf\def\PY@tc##1{\textcolor[rgb]{0.05,0.52,0.71}{##1}}}
\expandafter\def\csname PY@tok@nd\endcsname{\let\PY@bf=\textbf\def\PY@tc##1{\textcolor[rgb]{0.33,0.33,0.33}{##1}}}
\expandafter\def\csname PY@tok@ne\endcsname{\def\PY@tc##1{\textcolor[rgb]{0.00,0.44,0.13}{##1}}}
\expandafter\def\csname PY@tok@nf\endcsname{\def\PY@tc##1{\textcolor[rgb]{0.02,0.16,0.49}{##1}}}
\expandafter\def\csname PY@tok@si\endcsname{\let\PY@it=\textit\def\PY@tc##1{\textcolor[rgb]{0.44,0.63,0.82}{##1}}}
\expandafter\def\csname PY@tok@s2\endcsname{\def\PY@tc##1{\textcolor[rgb]{0.25,0.44,0.63}{##1}}}
\expandafter\def\csname PY@tok@nt\endcsname{\let\PY@bf=\textbf\def\PY@tc##1{\textcolor[rgb]{0.02,0.16,0.45}{##1}}}
\expandafter\def\csname PY@tok@nv\endcsname{\def\PY@tc##1{\textcolor[rgb]{0.73,0.38,0.84}{##1}}}
\expandafter\def\csname PY@tok@s1\endcsname{\def\PY@tc##1{\textcolor[rgb]{0.25,0.44,0.63}{##1}}}
\expandafter\def\csname PY@tok@ch\endcsname{\let\PY@it=\textit\def\PY@tc##1{\textcolor[rgb]{0.25,0.50,0.56}{##1}}}
\expandafter\def\csname PY@tok@m\endcsname{\def\PY@tc##1{\textcolor[rgb]{0.13,0.50,0.31}{##1}}}
\expandafter\def\csname PY@tok@gp\endcsname{\let\PY@bf=\textbf\def\PY@tc##1{\textcolor[rgb]{0.78,0.36,0.04}{##1}}}
\expandafter\def\csname PY@tok@sh\endcsname{\def\PY@tc##1{\textcolor[rgb]{0.25,0.44,0.63}{##1}}}
\expandafter\def\csname PY@tok@ow\endcsname{\let\PY@bf=\textbf\def\PY@tc##1{\textcolor[rgb]{0.00,0.44,0.13}{##1}}}
\expandafter\def\csname PY@tok@sx\endcsname{\def\PY@tc##1{\textcolor[rgb]{0.78,0.36,0.04}{##1}}}
\expandafter\def\csname PY@tok@bp\endcsname{\def\PY@tc##1{\textcolor[rgb]{0.00,0.44,0.13}{##1}}}
\expandafter\def\csname PY@tok@c1\endcsname{\let\PY@it=\textit\def\PY@tc##1{\textcolor[rgb]{0.25,0.50,0.56}{##1}}}
\expandafter\def\csname PY@tok@o\endcsname{\def\PY@tc##1{\textcolor[rgb]{0.40,0.40,0.40}{##1}}}
\expandafter\def\csname PY@tok@kc\endcsname{\let\PY@bf=\textbf\def\PY@tc##1{\textcolor[rgb]{0.00,0.44,0.13}{##1}}}
\expandafter\def\csname PY@tok@c\endcsname{\let\PY@it=\textit\def\PY@tc##1{\textcolor[rgb]{0.25,0.50,0.56}{##1}}}
\expandafter\def\csname PY@tok@mf\endcsname{\def\PY@tc##1{\textcolor[rgb]{0.13,0.50,0.31}{##1}}}
\expandafter\def\csname PY@tok@err\endcsname{\def\PY@bc##1{\setlength{\fboxsep}{0pt}\fcolorbox[rgb]{1.00,0.00,0.00}{1,1,1}{\strut ##1}}}
\expandafter\def\csname PY@tok@mb\endcsname{\def\PY@tc##1{\textcolor[rgb]{0.13,0.50,0.31}{##1}}}
\expandafter\def\csname PY@tok@ss\endcsname{\def\PY@tc##1{\textcolor[rgb]{0.32,0.47,0.09}{##1}}}
\expandafter\def\csname PY@tok@sr\endcsname{\def\PY@tc##1{\textcolor[rgb]{0.14,0.33,0.53}{##1}}}
\expandafter\def\csname PY@tok@mo\endcsname{\def\PY@tc##1{\textcolor[rgb]{0.13,0.50,0.31}{##1}}}
\expandafter\def\csname PY@tok@kd\endcsname{\let\PY@bf=\textbf\def\PY@tc##1{\textcolor[rgb]{0.00,0.44,0.13}{##1}}}
\expandafter\def\csname PY@tok@mi\endcsname{\def\PY@tc##1{\textcolor[rgb]{0.13,0.50,0.31}{##1}}}
\expandafter\def\csname PY@tok@kn\endcsname{\let\PY@bf=\textbf\def\PY@tc##1{\textcolor[rgb]{0.00,0.44,0.13}{##1}}}
\expandafter\def\csname PY@tok@cpf\endcsname{\let\PY@it=\textit\def\PY@tc##1{\textcolor[rgb]{0.25,0.50,0.56}{##1}}}
\expandafter\def\csname PY@tok@kr\endcsname{\let\PY@bf=\textbf\def\PY@tc##1{\textcolor[rgb]{0.00,0.44,0.13}{##1}}}
\expandafter\def\csname PY@tok@s\endcsname{\def\PY@tc##1{\textcolor[rgb]{0.25,0.44,0.63}{##1}}}
\expandafter\def\csname PY@tok@kp\endcsname{\def\PY@tc##1{\textcolor[rgb]{0.00,0.44,0.13}{##1}}}
\expandafter\def\csname PY@tok@w\endcsname{\def\PY@tc##1{\textcolor[rgb]{0.73,0.73,0.73}{##1}}}
\expandafter\def\csname PY@tok@kt\endcsname{\def\PY@tc##1{\textcolor[rgb]{0.56,0.13,0.00}{##1}}}
\expandafter\def\csname PY@tok@sc\endcsname{\def\PY@tc##1{\textcolor[rgb]{0.25,0.44,0.63}{##1}}}
\expandafter\def\csname PY@tok@sb\endcsname{\def\PY@tc##1{\textcolor[rgb]{0.25,0.44,0.63}{##1}}}
\expandafter\def\csname PY@tok@k\endcsname{\let\PY@bf=\textbf\def\PY@tc##1{\textcolor[rgb]{0.00,0.44,0.13}{##1}}}
\expandafter\def\csname PY@tok@se\endcsname{\let\PY@bf=\textbf\def\PY@tc##1{\textcolor[rgb]{0.25,0.44,0.63}{##1}}}
\expandafter\def\csname PY@tok@sd\endcsname{\let\PY@it=\textit\def\PY@tc##1{\textcolor[rgb]{0.25,0.44,0.63}{##1}}}

\def\PYZus{\char`\_}

\def\PYZsh{\char`\#}

\def\PYZhy{\char`\-}

\def\PYZdq{\char`\"}

\makeatother

\providecommand*{\DUrole}[2]{%
  \ifcsname DUrole#1\endcsname%
    \csname DUrole#1\endcsname{#2}%
  \else
    \ifcsname docutilsrole#1\endcsname%
      \csname docutilsrole#1\endcsname{#2}%
    \else%
      #2%
    \fi%
  \fi%
}

\providecommand*{\DUroletitlereference}[1]{\textsl{#1}}

\ifthenelse{\isundefined{\hypersetup}}{
  \usepackage[colorlinks=true,linkcolor=blue,urlcolor=blue]{hyperref}
  \urlstyle{same} 
}{}

\begin{document}
\newcounter{footnotecounter}\title{Massively parallel implementation in Python of a pseudo-spectral DNS code for turbulent flows}\author{Mikael Mortensen$^{\setcounter{footnotecounter}{1}\fnsymbol{footnotecounter}\setcounter{footnotecounter}{2}\fnsymbol{footnotecounter}}$%
          \setcounter{footnotecounter}{1}\thanks{\fnsymbol{footnotecounter} %
          Corresponding author: \protect\href{mailto:mikaem@math.uio.no}{mikaem@math.uio.no}}\setcounter{footnotecounter}{2}\thanks{\fnsymbol{footnotecounter} University of Oslo and Center for Biomedical Computing, Simula Research Laboratory}\thanks{%

          \noindent%
          Copyright\,\copyright\,2015 Mikael Mortensen. This is an open-access article distributed under the terms of the Creative Commons Attribution License, which permits unrestricted use, distribution, and reproduction in any medium, provided the original author and source are credited. http://creativecommons.org/licenses/by/3.0/%
        }}\maketitle
          \renewcommand{\leftmark}{PROC. OF THE 8th EUR. CONF. ON PYTHON IN SCIENCE (EUROSCIPY 2015)}
          \renewcommand{\rightmark}{MASSIVELY PARALLEL IMPLEMENTATION IN PYTHON OF A PSEUDO-SPECTRAL DNS CODE FOR TURBULENT FLOWS}

\InputIfFileExists{page_numbers.tex}{}{}
\newcommand*{\docutilsroleref}{\ref}
\newcommand*{\docutilsrolelabel}{\label}
\AtEndDocument{\cleardoublepage}
\begin{abstract}Direct Numerical Simulations (DNS) of the Navier Stokes equations is a
valuable research tool in fluid dynamics, but there are very few publicly
available codes and, due to heavy number crunching, codes are usually written
in low-level languages. In this work a \textasciitilde{}100 line standard scientific Python DNS code is described
that nearly matches the performance of pure C for thousands of processors
and billions of unknowns. With optimization of a few routines in Cython,
it is found to match the performance of a more or less identical solver
implemented from scratch in C++.

Keys to the efficiency of the solver are the mesh decomposition and three
dimensional FFT routines, implemented directly in Python using MPI, wrapped
through MPI for Python, and a serial FFT module (both \emph{numpy.fft} or \emph{pyFFTW} may be used).
Two popular decomposition strategies, \emph{slab} and \emph{pencil}, have been
implemented and tested.\end{abstract}

\section{Introduction%
  \label{introduction}%
}

Direct Numerical Simulations (DNS) of Navier Stokes equations have been used for decades to study fundamental aspects of turbulence and it is used extensively to validate turbulence models. DNS have been conducted on an extremely large scale on the largest supercomputers in the world. S. de Bruyn Kops \cite{deBruynKops} recently simulated homogeneous isotropic turbulence on a Cray XE6 architecture using a computational mesh with close to 1 trillion nodes ($8192^3$). Lee \emph{et al} \cite{Lee} simulated a turbulent channel flow on a Blue Gene/Q architecture using a mesh of size $15369 \times 1536 \times 11520$.

All known DNS codes (at least to the  knowledge of the author) running on supercomputers are implemented in low-level languages like Fortran or C/C++. These  languages are known for excellent performance in heavy duty number crunching algorithms, which goes a long way to explain the popularity. Python, on the other hand, is a scripting language known for being very convenient to work with, but as a research tool more aimed at post-processing, visualization or fast prototyping than high performance computing. However, a lesser known fact is that Python is very convenient to program also with MPI, and that as long as number crunching is performed using vectorized expressions, a code may run on thousands of processors at speeds closing in on the optimal low-level codes.

The purpose of this work is to describe a \textasciitilde{}100 line pseudo-spectral DNS solver developed from scratch in Python, using nothing more than NumPy and MPI for Python (mpi4py), possibly optimized with pyFFTW and Cython. It is important to stress that the entire solver is written in Python, this is not simply a wrapper of a low-level number cruncher. The mesh is created and decomposed in Python and MPI communications are implemented using mpi4py. Two popular strategies, \emph{slab} and \emph{pencil}, for MPI communications of the three-dimensional Fast Fourier Transform (FFT), required by the pseudo-spectral method, will be described. The entire solver is available online (\url{https://github.com/mikaem/spectralDNS}) under the GPL license.

In this short paper we will first describe the Fourier transformed Navier Stokes equations that are solved for a triply periodic domain. We will then give a brief description of the implementation and show the results of performance tests conducted on a BlueGene/P supercomputer at the KAUST supercomputing laboratory. The performance of the scientific Python solver, as well as a version optimized with Cython, is compared to a pure C++ implementation.

\section{Navier Stokes in Fourier space%
  \label{navier-stokes-in-fourier-space}%
}

Turbulent flows are described by the Navier Stokes equations. DNS of the Navier-Stokes equations are often performed in periodic domains to allow the study of pure isotropic turbulence and to avoid inhomogeneities associated with flows near walls. The periodicity of the solution also allows us to lift the equations to Fourier space and to use highly accurate Fourier spectral discretization of space. In this work we consider a triply periodic domain and we use a spectral Fourier-Galerkin method \cite{canuto1988} for the spatial discretization. To arrive at the equations being solved we first cast the Navier-Stokes equations in rotational form\begin{eqnarray}
\label{eqNS}
\frac{\partial \bm{u}}{\partial t} - \bm{u} \times \bm{\omega}   &=& \nu \nabla^2 \bm{u} - \nabla{P}, \\
\nabla \cdot \bm{u} &=& 0, \\
\bm{u}(\bm{x}+2\pi \bm{e}^i, t) &=& \bm{u}(\bm{x}, t), \quad \text{for }\, i=1,2,3,\\
\bm{u}(\bm{x}, 0) &=& \bm{u}_0(\bm{x})
\end{eqnarray}where $\bm{u}(\bm{x}, t)$ is the velocity vector, $\bm{\omega}=\nabla \times \bm{u}$ the vorticity vector, $\bm{e}^i$ the Cartesian unit vectors, and the modified pressure $P=p+\bm{u}\cdot \bm{u}/2$, where $p$ is the regular pressure normalized by the constant density. The equations are periodic in all three spatial directions. If all three directions now are discretized uniformly in space using a structured computational mesh with $N$ points in each direction, the mesh, $\bm{x}=(x,y,z)$, can be represented as\begin{equation}
\label{eq:realmesh}
\bm{x} =(x_i, y_j, z_k) = \left\{\left( \frac{2\pi i}{N}, \frac{2\pi j}{N}, \frac{2\pi k}{N} \right): i,j,k \in 0,\ldots, N-1\right\} .
\end{equation}To transform the equations from real space to Fourier space we need the corresponding wavenumber mesh\begin{equation}
\label{eq:wavemesh}
\bm{k} = (k_x, k_y, k_z) = \left\{(l, m, n): \, l, m, n \in -\frac{N}{2}+1,\ldots, \frac{N}{2} \right\},
\end{equation}and to move back and forth between real and wavenumber space we use the three-dimensional Fourier transforms\begin{eqnarray}
\label{eq:ffteq}
u(\bm{x}, t) &=& \frac{1}{N^3}\sum_{\bm{k}} \hat{u}_{\bm{k}}(t) e^{\imath \bm{k}\cdot \bm{x}}, \\
\hat{u}_{\bm{k}}(t) &=& \sum_{\bm{x}} u(\bm{x}, t) e^{-\imath \bm{k}\cdot \bm{x}}
\end{eqnarray}where $\hat{u}_{\bm{k}}(t)$ is used to represent the Fourier coefficients and $\imath=\sqrt{-1}$ represents the imaginary unit. The exponential $e^{\imath \bm{k}\cdot \bm{x}}$ represents the basis functions for the spectral Fourier-Galerkin method. To simplify we use the notation\begin{eqnarray*}
\hat{u}_{\bm{k}}(t) &=& \mathcal{F}({u}(\bm{x}, t)) \left[= \mathcal{F}_{k_x} \left(\mathcal{F}_{k_y} \left( \mathcal{F}_{k_z} ({u}) \right) \right) \right], \\
{u}(\bm{x}, t) &=& \mathcal{F}^{-1}(\hat{u}_{\bm{k}}(t)) \left[= \mathcal{F}^{-1}_{z}\left(\mathcal{F}^{-1}_{y}\left(\mathcal{F}^{-1}_{x}(\hat{{u}})\right)\right)\right],
\end{eqnarray*}where the forward and inverse Fourier transforms are, respectively, $\mathcal{F}$ and $\mathcal{F}^{-1}$. The square bracket shows the direction of the three consecutive transforms in three-dimensional space. The order of the directions are irrelevant, but the inverse needs to be in the opposite order of the forward transform.

In the spectral Fourier-Galerkin method it is possible to reduce the set of four partial differential equations (\DUrole{ref}{eqNS}) to three ordinary differential equations. To this end Eq. (\DUrole{ref}{eqNS}) is first transformed by multiplying with the test function $e^{-\imath \bm{k}\cdot \bm{x}}$ and integrating over the domain. The pressure may then be eliminated by dotting this transformed equation by $\imath \bm{k}$ and using the divergence constraint (in spectral space $\nabla \cdot \bm{u} = \imath \bm{k}\cdot \bm{u}_{\bm{k}}$). The eact equation for the pressure then reads\begin{equation}
\label{eq:pressure}
\hat{P}_{\bm{k}} = - \frac{\imath\bm{k} \cdot \widehat{( \bm{u} \times \bm{\omega})}_{\bm{k}} }{|\bm{k}|^2},
\end{equation}and this is used to eliminate the pressure from the momentum equation. We finally obtain ordinary differential equations for the three transformed velocity components\begin{equation}
\label{eq:NSfinal}
\frac{d\hat{\bm{u}}_{\bm{k}}}{d t}  = \widehat{( \bm{u} \times \bm{\omega})}_{\bm{k}} - \nu |\bm{k}|^2  \hat{\bm{u}}_{\bm{k}} - \bm{k} \frac{\bm{k} \cdot \widehat{( \bm{u} \times \bm{\omega})}_{\bm{k}} }{|\bm{k}|^2}.
\end{equation}An explicit solver will integrate Eq. \DUrole{ref}{eq:NSfinal} from given initial conditions. Any integrator may be used, here we have settled for a fourth order \cite{Runge-Kutta} method.

\section{Details of implementation%
  \label{details-of-implementation}%
}

The major challenges one has to deal with when implementing a high performance solver for Eq. (\DUrole{ref}{eq:NSfinal}) in Python are the following%
\begin{itemize}

\item 

MPI
\item 

Mesh decomposition
\item 

Three dimensional Fourier transforms with MPI
\item 

Vectorization (NumPy ufuncs)
\item 

Dynamic loading of Python on a supercomputer
\end{itemize}

\subsection{MPI/MPI for Python (mpi4py)%
  \label{mpi-mpi-for-python-mpi4py}%
}

The \cite{mpi4py} Python package contains wrappers for almost the entire MPI and it has been shown to be able to distribute NumPy arrays at the speed of regular C arrays. The MPI for Python module allows us to write Python code with MPI just like regular low-level languages, but with a much simpler and user-friendly syntax. Since coding is performed like in C, the Python implementation may, as such, be used as an easy to follow, working prototype for a complete low-level implementation in Fortran, C or C++.

\subsection{Mesh decomposition%
  \label{mesh-decomposition}%
}

The computational mesh is structured and the most common approaches to mesh decomposition are the \emph{slab} and the \emph{pencil} methods. The \emph{slab} decomposition distributes the mesh along one single index, whereas the \emph{pencil} distributes two of the three indices. The advantage of the \emph{slab} decomposition is that it is generally faster than \emph{pencil}, but it is limited to $N$ CPUs for a computational mesh of size $N^3$. The \emph{pencil} decomposition is slower, but has the advantage that it can be used by $N^2$ CPUs and thus allows for much larger simulations. Figure \DUrole{ref}{slab} shows how the distributed mesh is laid out for \emph{slab} decomposition using 4 CPUs. Notice that in real space the decomposition is along the first index, whereas in wavenumber space it is along the second index. This is because the third and final FFT is performed along the x-direction, and for this operation the mesh needs to be aligned either in the x-z plane or in the x-y plane. Here we have simply chosen the first option.\begin{figure}[bht]\noindent\makebox[\columnwidth][c]{\includegraphics[scale=0.15]{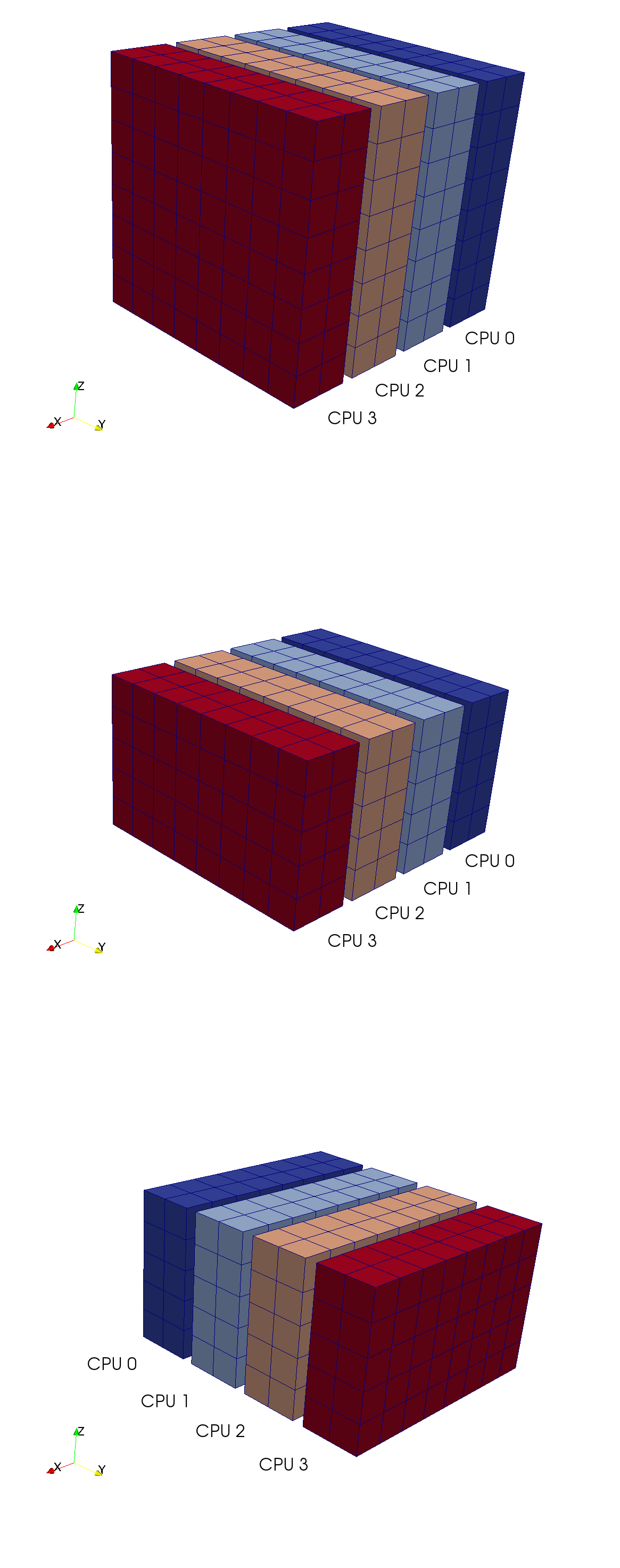}}
\caption{From top to bottom slab decomposition of physical mesh, intermediate wavenumber mesh and final wavenumber mesh respectively. \DUrole{label}{slab}}
\end{figure}

\subsection{Three dimensional Fourier transforms with MPI%
  \label{three-dimensional-fourier-transforms-with-mpi}%
}

The regular Python modules \DUroletitlereference{numpy.fft}, \DUroletitlereference{scipy.fftpack} and \cite{pyfftw} all provide routines to do FFTs on regular (non-distributed) structured meshes along any given axis. Any one of these modules may be used, and the only challenge is that the FFTs need to be performed in parallel with MPI. None of the regular Python modules have routines to do FFT in parallel, and the main reason for this is that the FFTs need to be performed on a distributed mesh, where the mesh is distributed before the FFT routines are called. In this work we present 3D FFT routines with MPI for both the \emph{slab} and the \emph{pencil} decomposition. The FFTs themselves are performed on data local to one single processor, and hence the serial FFT of any provider may be used. All other operations required to perform the 3D FFT are implemented in Python. This includes both transpose operations and an MPI call to the \DUroletitlereference{Alltoall} function. The entire Python implementation of the 3D FFT with MPI for a \emph{slab} mesh is shown below\begin{Verbatim}[commandchars=\\\{\},fontsize=\footnotesize]
\PY{k+kn}{from} \PY{n+nn}{pyfftw} \PY{k+kn}{import} \PY{n}{fft}\PY{p}{,} \PY{n}{ifft}\PY{p}{,} \PY{n}{rfft2}\PY{p}{,} \PY{n}{irfft2}\PY{p}{,} \PY{n}{empty}

\PY{c+c1}{\PYZsh{} Preallocated work array for MPI}
\PY{n}{U\PYZus{}mpi} \PY{o}{=} \PY{n}{empty}\PY{p}{(}\PY{p}{(}\PY{n}{num\PYZus{}processes}\PY{p}{,} \PY{n}{Np}\PY{p}{,} \PY{n}{Np}\PY{p}{,} \PY{n}{Nf}\PY{p}{)}\PY{p}{,}
              \PY{n}{dtype}\PY{o}{=}\PY{n+nb}{complex}\PY{p}{)}

\PY{k}{def} \PY{n+nf}{fftn\PYZus{}mpi}\PY{p}{(}\PY{n}{u}\PY{p}{,} \PY{n}{fu}\PY{p}{)}\PY{p}{:}
    \PY{l+s+sd}{\PYZdq{}\PYZdq{}\PYZdq{}FFT in three directions using MPI.\PYZdq{}\PYZdq{}\PYZdq{}}
    \PY{n}{Uc\PYZus{}hatT} \PY{o}{=} \PY{n}{rfft2}\PY{p}{(}\PY{n}{u}\PY{p}{,} \PY{n}{axes}\PY{o}{=}\PY{p}{(}\PY{l+m+mi}{1}\PY{p}{,}\PY{l+m+mi}{2}\PY{p}{)}\PY{p}{)}
    \PY{k}{for} \PY{n}{i} \PY{o+ow}{in} \PY{n+nb}{range}\PY{p}{(}\PY{n}{num\PYZus{}processes}\PY{p}{)}\PY{p}{:}
        \PY{n}{U\PYZus{}mpi}\PY{p}{[}\PY{n}{i}\PY{p}{]} \PY{o}{=} \PY{n}{Uc\PYZus{}hatT}\PY{p}{[}\PY{p}{:}\PY{p}{,} \PY{n}{i}\PY{o}{*}\PY{n}{Np}\PY{p}{:}\PY{p}{(}\PY{n}{i}\PY{o}{+}\PY{l+m+mi}{1}\PY{p}{)}\PY{o}{*}\PY{n}{Np}\PY{p}{]}
    \PY{n}{comm}\PY{o}{.}\PY{n}{Alltoall}\PY{p}{(}\PY{p}{[}\PY{n}{U\PYZus{}mpi}\PY{p}{,} \PY{n}{mpitype}\PY{p}{]}\PY{p}{,} \PY{p}{[}\PY{n}{fu}\PY{p}{,} \PY{n}{mpitype}\PY{p}{]}\PY{p}{)}
    \PY{n}{fu} \PY{o}{=} \PY{n}{fft}\PY{p}{(}\PY{n}{fu}\PY{p}{,} \PY{n}{axis}\PY{o}{=}\PY{l+m+mi}{0}\PY{p}{)}
    \PY{k}{return} \PY{n}{fu}

\PY{k}{def} \PY{n+nf}{ifftn\PYZus{}mpi}\PY{p}{(}\PY{n}{fu}\PY{p}{,} \PY{n}{u}\PY{p}{)}\PY{p}{:}
    \PY{l+s+sd}{\PYZdq{}\PYZdq{}\PYZdq{}Inverse FFT in three directions using MPI.}
\PY{l+s+sd}{       Need to do ifft in reversed order of fft.\PYZdq{}\PYZdq{}\PYZdq{}}
    \PY{n}{Uc\PYZus{}hat} \PY{o}{=} \PY{n}{ifft}\PY{p}{(}\PY{n}{fu}\PY{p}{,} \PY{n}{axis}\PY{o}{=}\PY{l+m+mi}{0}\PY{p}{)}
    \PY{n}{comm}\PY{o}{.}\PY{n}{Alltoall}\PY{p}{(}\PY{p}{[}\PY{n}{Uc\PYZus{}hat}\PY{p}{,} \PY{n}{mpitype}\PY{p}{]}\PY{p}{,} \PY{p}{[}\PY{n}{U\PYZus{}mpi}\PY{p}{,} \PY{n}{mpitype}\PY{p}{]}\PY{p}{)}
    \PY{k}{for} \PY{n}{i} \PY{o+ow}{in} \PY{n+nb}{range}\PY{p}{(}\PY{n}{num\PYZus{}processes}\PY{p}{)}\PY{p}{:}
        \PY{n}{Uc\PYZus{}hatT}\PY{p}{[}\PY{p}{:}\PY{p}{,} \PY{p}{:}\PY{p}{,} \PY{n}{i}\PY{o}{*}\PY{n}{Np}\PY{p}{:}\PY{p}{(}\PY{n}{i}\PY{o}{+}\PY{l+m+mi}{1}\PY{p}{)}\PY{o}{*}\PY{n}{Np}\PY{p}{]} \PY{o}{=} \PY{n}{U\PYZus{}mpi}\PY{p}{[}\PY{n}{i}\PY{p}{]}
    \PY{n}{u} \PY{o}{=} \PY{n}{irfft2}\PY{p}{(}\PY{n}{Uc\PYZus{}hatT}\PY{p}{,} \PY{n}{axes}\PY{o}{=}\PY{p}{(}\PY{l+m+mi}{2}\PY{p}{,}\PY{l+m+mi}{1}\PY{p}{)}\PY{p}{)}
    \PY{k}{return} \PY{n}{u}
\end{Verbatim}
Note that merely one single work array needs to be pre-allocated for the collective call to \DUroletitlereference{Alltoall}. The \DUroletitlereference{pyFFTW} wrapping of the \DUroletitlereference{libFFTW} library allocates internally work arrays for both input and output arrays, and the pointers \DUroletitlereference{Uc\_hatT} and \DUroletitlereference{Uc\_hat} above are simply references to this internal storage.

For short of space the implementation for the \emph{pencil} decomposition is not shown here, but it requires about twice the amount of code since the mesh needs to be transformed and distributed twice (along two indices).

\subsection{Vectorization and NumPy ufuncs%
  \label{vectorization-and-numpy-ufuncs}%
}

Besides the FFTs, the major computational cost of the pseudo-spectral solver lies in element-wise multiplications, divisions, subtractions and additions that are required to assemble the right hand side of Eq (\DUrole{ref}{eq:NSfinal}). For efficiency it is imperative that the NumPy code is vectorized, thus avoiding for-loops that are very expensive in Python. When properly vectorized the element-wise operations are carried out by NumPy universal functions (so called ufuncs), calling compiled C-code on loops over the entire (or parts of) the data structures. When properly set up many arithmetic operations may be performed at near optimal speed, but, unfortunately, complex expressions are known to be rather slow compared to low-level implementations due to multiple calls to the same loop and the creation of temporary arrays. The \cite{numexpr} module has actually been created with the specific goal of speeding up such element-wise complex expressions. Besides \DUroletitlereference{numexpr}, the most common ways of speeding up scientific Python code is through \cite{Cython}, \cite{Numba} or \cite{weave}.

Two bottlenecks appear in the standard scientific Python implementation of the pseudo spectral solver. The first is the \emph{for} loops seen in the \emph{fftn\_mpi/ifftn\_mpi} functions previously described. The second is the cross product that needs to be computed in Eq. (\DUrole{ref}{eq:NSfinal}). A straight forward vectorized implementation and usage of the cross product is\begin{Verbatim}[commandchars=\\\{\},fontsize=\footnotesize]
\PY{k+kn}{import} \PY{n+nn}{numpy}

\PY{k}{def} \PY{n+nf}{cross}\PY{p}{(}\PY{n}{c}\PY{p}{,} \PY{n}{a}\PY{p}{,} \PY{n}{b}\PY{p}{)}\PY{p}{:}
    \PY{l+s+sd}{\PYZdq{}\PYZdq{}\PYZdq{}Regular c = a x b\PYZdq{}\PYZdq{}\PYZdq{}}
    \PY{c+c1}{\PYZsh{}c[:] = numpy.cross(a, b, axis=0)}
    \PY{n}{c}\PY{p}{[}\PY{l+m+mi}{0}\PY{p}{]} \PY{o}{=} \PY{n}{a}\PY{p}{[}\PY{l+m+mi}{1}\PY{p}{]}\PY{o}{*}\PY{n}{b}\PY{p}{[}\PY{l+m+mi}{2}\PY{p}{]} \PY{o}{\PYZhy{}} \PY{n}{a}\PY{p}{[}\PY{l+m+mi}{2}\PY{p}{]}\PY{o}{*}\PY{n}{b}\PY{p}{[}\PY{l+m+mi}{1}\PY{p}{]}
    \PY{n}{c}\PY{p}{[}\PY{l+m+mi}{1}\PY{p}{]} \PY{o}{=} \PY{n}{a}\PY{p}{[}\PY{l+m+mi}{2}\PY{p}{]}\PY{o}{*}\PY{n}{b}\PY{p}{[}\PY{l+m+mi}{0}\PY{p}{]} \PY{o}{\PYZhy{}} \PY{n}{a}\PY{p}{[}\PY{l+m+mi}{0}\PY{p}{]}\PY{o}{*}\PY{n}{b}\PY{p}{[}\PY{l+m+mi}{2}\PY{p}{]}
    \PY{n}{c}\PY{p}{[}\PY{l+m+mi}{2}\PY{p}{]} \PY{o}{=} \PY{n}{a}\PY{p}{[}\PY{l+m+mi}{0}\PY{p}{]}\PY{o}{*}\PY{n}{b}\PY{p}{[}\PY{l+m+mi}{1}\PY{p}{]} \PY{o}{\PYZhy{}} \PY{n}{a}\PY{p}{[}\PY{l+m+mi}{1}\PY{p}{]}\PY{o}{*}\PY{n}{b}\PY{p}{[}\PY{l+m+mi}{0}\PY{p}{]}
    \PY{k}{return} \PY{n}{c}

\PY{c+c1}{\PYZsh{} Usage}
\PY{n}{N} \PY{o}{=} \PY{l+m+mi}{200}
\PY{n}{U} \PY{o}{=} \PY{n}{numpy}\PY{o}{.}\PY{n}{zeros}\PY{p}{(}\PY{p}{(}\PY{l+m+mi}{3}\PY{p}{,} \PY{n}{N}\PY{p}{,} \PY{n}{N}\PY{p}{,} \PY{n}{N}\PY{p}{)}\PY{p}{)}
\PY{n}{W} \PY{o}{=} \PY{n}{numpy}\PY{o}{.}\PY{n}{zeros}\PY{p}{(}\PY{p}{(}\PY{l+m+mi}{3}\PY{p}{,} \PY{n}{N}\PY{p}{,} \PY{n}{N}\PY{p}{,} \PY{n}{N}\PY{p}{)}\PY{p}{)}
\PY{n}{F} \PY{o}{=} \PY{n}{numpyzeros}\PY{p}{(}\PY{p}{(}\PY{l+m+mi}{3}\PY{p}{,} \PY{n}{N}\PY{p}{,} \PY{n}{N}\PY{p}{,} \PY{n}{N}\PY{p}{)}\PY{p}{)}
\PY{n}{F} \PY{o}{=} \PY{n}{cross}\PY{p}{(}\PY{n}{U}\PY{p}{,} \PY{n}{W}\PY{p}{,} \PY{n}{F}\PY{p}{)}
\end{Verbatim}
The cross product actually makes 6 calls to the multiply ufunc, 3 to subtract, and also requires temporary arrays for storage. Each ufunc loops over the entire computational mesh and as such it is not unexpected that the computation of the cross product becomes a bottleneck. The built-in \DUroletitlereference{numpy.cross} (shown in the cross code listing) uses ufuncs as well and runs approximately as fast as the code shown. Moving this routine to Numba or Cython we can hardcode the loop over the mesh just once and speed-up is approximately a factor of 5. A Numba implementation is shown below\begin{Verbatim}[commandchars=\\\{\},fontsize=\footnotesize]
\PY{k+kn}{from} \PY{n+nn}{numba} \PY{k+kn}{import} \PY{n}{jit}\PY{p}{,} \PY{n}{float64} \PY{k}{as} \PY{n+nb}{float}

\PY{n+nd}{@jit}\PY{p}{(}\PY{n+nb}{float}\PY{p}{[}\PY{p}{:}\PY{p}{,}\PY{p}{:}\PY{p}{,}\PY{p}{:}\PY{p}{,}\PY{p}{:}\PY{p}{]}\PY{p}{(}\PY{n+nb}{float}\PY{p}{[}\PY{p}{:}\PY{p}{,}\PY{p}{:}\PY{p}{,}\PY{p}{:}\PY{p}{,}\PY{p}{:}\PY{p}{]}\PY{p}{,}
     \PY{n+nb}{float}\PY{p}{[}\PY{p}{:}\PY{p}{,}\PY{p}{:}\PY{p}{,}\PY{p}{:}\PY{p}{,}\PY{p}{:}\PY{p}{]}\PY{p}{,} \PY{n+nb}{float}\PY{p}{[}\PY{p}{:}\PY{p}{,}\PY{p}{:}\PY{p}{,}\PY{p}{:}\PY{p}{,}\PY{p}{:}\PY{p}{]}\PY{p}{)}\PY{p}{,} \PY{n}{nopython}\PY{o}{=}\PY{n+nb+bp}{True}\PY{p}{)}
\PY{k}{def} \PY{n+nf}{cross}\PY{p}{(}\PY{n}{a}\PY{p}{,} \PY{n}{b}\PY{p}{,} \PY{n}{c}\PY{p}{)}\PY{p}{:}
    \PY{k}{for} \PY{n}{i} \PY{o+ow}{in} \PY{n+nb}{xrange}\PY{p}{(}\PY{n}{a}\PY{o}{.}\PY{n}{shape}\PY{p}{[}\PY{l+m+mi}{1}\PY{p}{]}\PY{p}{)}\PY{p}{:}
        \PY{k}{for} \PY{n}{j} \PY{o+ow}{in} \PY{n+nb}{xrange}\PY{p}{(}\PY{n}{a}\PY{o}{.}\PY{n}{shape}\PY{p}{[}\PY{l+m+mi}{2}\PY{p}{]}\PY{p}{)}\PY{p}{:}
            \PY{k}{for} \PY{n}{k} \PY{o+ow}{in} \PY{n+nb}{xrange}\PY{p}{(}\PY{n}{a}\PY{o}{.}\PY{n}{shape}\PY{p}{[}\PY{l+m+mi}{3}\PY{p}{]}\PY{p}{)}\PY{p}{:}
                \PY{n}{a0} \PY{o}{=} \PY{n}{a}\PY{p}{[}\PY{l+m+mi}{0}\PY{p}{,}\PY{n}{i}\PY{p}{,}\PY{n}{j}\PY{p}{,}\PY{n}{k}\PY{p}{]}
                \PY{n}{a1} \PY{o}{=} \PY{n}{a}\PY{p}{[}\PY{l+m+mi}{1}\PY{p}{,}\PY{n}{i}\PY{p}{,}\PY{n}{j}\PY{p}{,}\PY{n}{k}\PY{p}{]}
                \PY{n}{a2} \PY{o}{=} \PY{n}{a}\PY{p}{[}\PY{l+m+mi}{2}\PY{p}{,}\PY{n}{i}\PY{p}{,}\PY{n}{j}\PY{p}{,}\PY{n}{k}\PY{p}{]}
                \PY{n}{b0} \PY{o}{=} \PY{n}{b}\PY{p}{[}\PY{l+m+mi}{0}\PY{p}{,}\PY{n}{i}\PY{p}{,}\PY{n}{j}\PY{p}{,}\PY{n}{k}\PY{p}{]}
                \PY{n}{b1} \PY{o}{=} \PY{n}{b}\PY{p}{[}\PY{l+m+mi}{1}\PY{p}{,}\PY{n}{i}\PY{p}{,}\PY{n}{j}\PY{p}{,}\PY{n}{k}\PY{p}{]}
                \PY{n}{b2} \PY{o}{=} \PY{n}{b}\PY{p}{[}\PY{l+m+mi}{2}\PY{p}{,}\PY{n}{i}\PY{p}{,}\PY{n}{j}\PY{p}{,}\PY{n}{k}\PY{p}{]}
                \PY{n}{c}\PY{p}{[}\PY{l+m+mi}{0}\PY{p}{,}\PY{n}{i}\PY{p}{,}\PY{n}{j}\PY{p}{,}\PY{n}{k}\PY{p}{]} \PY{o}{=} \PY{n}{a1}\PY{o}{*}\PY{n}{b2} \PY{o}{\PYZhy{}} \PY{n}{a2}\PY{o}{*}\PY{n}{b1}
                \PY{n}{c}\PY{p}{[}\PY{l+m+mi}{1}\PY{p}{,}\PY{n}{i}\PY{p}{,}\PY{n}{j}\PY{p}{,}\PY{n}{k}\PY{p}{]} \PY{o}{=} \PY{n}{a2}\PY{o}{*}\PY{n}{b0} \PY{o}{\PYZhy{}} \PY{n}{a0}\PY{o}{*}\PY{n}{b2}
                \PY{n}{c}\PY{p}{[}\PY{l+m+mi}{2}\PY{p}{,}\PY{n}{i}\PY{p}{,}\PY{n}{j}\PY{p}{,}\PY{n}{k}\PY{p}{]} \PY{o}{=} \PY{n}{a0}\PY{o}{*}\PY{n}{b1} \PY{o}{\PYZhy{}} \PY{n}{a1}\PY{o}{*}\PY{n}{b0}
    \PY{k}{return} \PY{n}{c}
\end{Verbatim}
The Numba code works out of the box and is compiled on the fly by a just-in-time compiler. A Cython version looks very similar, but requires compilation into a module that is subsequently imported back into python. The Cython code below uses fused types to generate code for single and double precision simultaneously.\begin{Verbatim}[commandchars=\\\{\},fontsize=\footnotesize]
\PY{n}{cimport} \PY{n}{numpy} \PY{k}{as} \PY{n}{np}
\PY{n}{ctypedef} \PY{n}{fused} \PY{n}{T}\PY{p}{:}
    \PY{n}{np}\PY{o}{.}\PY{n}{float64\PYZus{}t}
    \PY{n}{np}\PY{o}{.}\PY{n}{float32\PYZus{}t}

\PY{k}{def} \PY{n+nf}{cross}\PY{p}{(}\PY{n}{np}\PY{o}{.}\PY{n}{ndarray}\PY{p}{[}\PY{n}{T}\PY{p}{,} \PY{n}{ndim}\PY{o}{=}\PY{l+m+mi}{4}\PY{p}{]} \PY{n}{a}\PY{p}{,}
          \PY{n}{np}\PY{o}{.}\PY{n}{ndarray}\PY{p}{[}\PY{n}{T}\PY{p}{,} \PY{n}{ndim}\PY{o}{=}\PY{l+m+mi}{4}\PY{p}{]} \PY{n}{b}\PY{p}{,}
          \PY{n}{np}\PY{o}{.}\PY{n}{ndarray}\PY{p}{[}\PY{n}{T}\PY{p}{,} \PY{n}{ndim}\PY{o}{=}\PY{l+m+mi}{4}\PY{p}{]} \PY{n}{c}\PY{p}{)}\PY{p}{:}
    \PY{n}{cdef} \PY{n}{unsigned} \PY{n+nb}{int} \PY{n}{i}\PY{p}{,} \PY{n}{j}\PY{p}{,} \PY{n}{k}
    \PY{n}{cdef} \PY{n}{T} \PY{n}{a0}\PY{p}{,} \PY{n}{a1}\PY{p}{,} \PY{n}{a2}\PY{p}{,} \PY{n}{b0}\PY{p}{,} \PY{n}{b1}\PY{p}{,} \PY{n}{b2}
    \PY{k}{for} \PY{n}{i} \PY{o+ow}{in} \PY{n+nb}{xrange}\PY{p}{(}\PY{n}{a}\PY{o}{.}\PY{n}{shape}\PY{p}{[}\PY{l+m+mi}{1}\PY{p}{]}\PY{p}{)}\PY{p}{:}
        \PY{k}{for} \PY{n}{j} \PY{o+ow}{in} \PY{n+nb}{xrange}\PY{p}{(}\PY{n}{a}\PY{o}{.}\PY{n}{shape}\PY{p}{[}\PY{l+m+mi}{2}\PY{p}{]}\PY{p}{)}\PY{p}{:}
            \PY{k}{for} \PY{n}{k} \PY{o+ow}{in} \PY{n+nb}{xrange}\PY{p}{(}\PY{n}{a}\PY{o}{.}\PY{n}{shape}\PY{p}{[}\PY{l+m+mi}{3}\PY{p}{]}\PY{p}{)}\PY{p}{:}
                \PY{n}{a0} \PY{o}{=} \PY{n}{a}\PY{p}{[}\PY{l+m+mi}{0}\PY{p}{,}\PY{n}{i}\PY{p}{,}\PY{n}{j}\PY{p}{,}\PY{n}{k}\PY{p}{]}
                \PY{n}{a1} \PY{o}{=} \PY{n}{a}\PY{p}{[}\PY{l+m+mi}{1}\PY{p}{,}\PY{n}{i}\PY{p}{,}\PY{n}{j}\PY{p}{,}\PY{n}{k}\PY{p}{]}
                \PY{n}{a2} \PY{o}{=} \PY{n}{a}\PY{p}{[}\PY{l+m+mi}{2}\PY{p}{,}\PY{n}{i}\PY{p}{,}\PY{n}{j}\PY{p}{,}\PY{n}{k}\PY{p}{]}
                \PY{n}{b0} \PY{o}{=} \PY{n}{b}\PY{p}{[}\PY{l+m+mi}{0}\PY{p}{,}\PY{n}{i}\PY{p}{,}\PY{n}{j}\PY{p}{,}\PY{n}{k}\PY{p}{]}
                \PY{n}{b1} \PY{o}{=} \PY{n}{b}\PY{p}{[}\PY{l+m+mi}{1}\PY{p}{,}\PY{n}{i}\PY{p}{,}\PY{n}{j}\PY{p}{,}\PY{n}{k}\PY{p}{]}
                \PY{n}{b2} \PY{o}{=} \PY{n}{b}\PY{p}{[}\PY{l+m+mi}{2}\PY{p}{,}\PY{n}{i}\PY{p}{,}\PY{n}{j}\PY{p}{,}\PY{n}{k}\PY{p}{]}
                \PY{n}{c}\PY{p}{[}\PY{l+m+mi}{0}\PY{p}{,}\PY{n}{i}\PY{p}{,}\PY{n}{j}\PY{p}{,}\PY{n}{k}\PY{p}{]} \PY{o}{=} \PY{n}{a1}\PY{o}{*}\PY{n}{b2} \PY{o}{\PYZhy{}} \PY{n}{a2}\PY{o}{*}\PY{n}{b1}
                \PY{n}{c}\PY{p}{[}\PY{l+m+mi}{1}\PY{p}{,}\PY{n}{i}\PY{p}{,}\PY{n}{j}\PY{p}{,}\PY{n}{k}\PY{p}{]} \PY{o}{=} \PY{n}{a2}\PY{o}{*}\PY{n}{b0} \PY{o}{\PYZhy{}} \PY{n}{a0}\PY{o}{*}\PY{n}{b2}
                \PY{n}{c}\PY{p}{[}\PY{l+m+mi}{2}\PY{p}{,}\PY{n}{i}\PY{p}{,}\PY{n}{j}\PY{p}{,}\PY{n}{k}\PY{p}{]} \PY{o}{=} \PY{n}{a0}\PY{o}{*}\PY{n}{b1} \PY{o}{\PYZhy{}} \PY{n}{a1}\PY{o}{*}\PY{n}{b0}
    \PY{k}{return} \PY{n}{c}
\end{Verbatim}
In addition, both \emph{scipy.weave} and \emph{numexpr} have been tested as well, but they have been found to be slower than Numba and Cython.

\subsection{Dynamic loading of Python on supercomputers%
  \label{dynamic-loading-of-python-on-supercomputers}%
}

The dynamic loading of Python on supercomputers can be very slow due to bottlenecks in the filesystem when thousands of processors attempt to open the same files. A solution to this problem has been provided by the scalable Python version developed by J. \cite{Enkovaara} and used by \cite{GPAW}, where CPython is modified slightly such that during import operations only a single process performs the actual I/O, and MPI is used for broadcasting the data to other MPI ranks. With scalable Python the dynamic loading times are kept at approximately 30 seconds for a full rack (4096 cores).

\section{Parallel scaling on Blue Gene/P%
  \label{parallel-scaling-on-blue-gene-p}%
}

In this section we compare the performance of the solver with a pure C++ implementation on Shaheen, a Blue Gene/P supercomputer at the KAUST supercomputing Laboratory. The C++ solver we are comparing with has been implemented using the Python solver as prototype and the only real difference is that the C++ solver is using the 3D FFT routines from \cite{FFTW} with MPI included. For optimization we are only considering the Cython implementation, because we were not able to install Numba on Shaheen.

The solver is run for a Taylor Green test case initialized as\begin{eqnarray}
\label{TG}
 u(x, y, z) &=& \sin(x)  \cos(y) \cos(z), \notag \\
 {v}(x, y, z) &=&-\cos(x) \sin(y) \cos(z), \notag\\
 {w}(x, y, z) &=& 0, \notag
\end{eqnarray}with a Reynolds number of 1600 and a time step of 0.001. At first the implementation is verified by running the solver for a time $t=[0, 20]$ and comparing the results to a previously verified reference solution, generated from a well tested and established low-level pseudo-spectral solver and utilized by the annual International Workshop on High-Order \cite{CFD} Methods. From start to finish, over 20,000 time steps, the L2 error norm of the solution computed by our solver never strays more than 1e-6 from the reference solution.\begin{figure}[bht]\noindent\makebox[\columnwidth][c]{\includegraphics[scale=0.50]{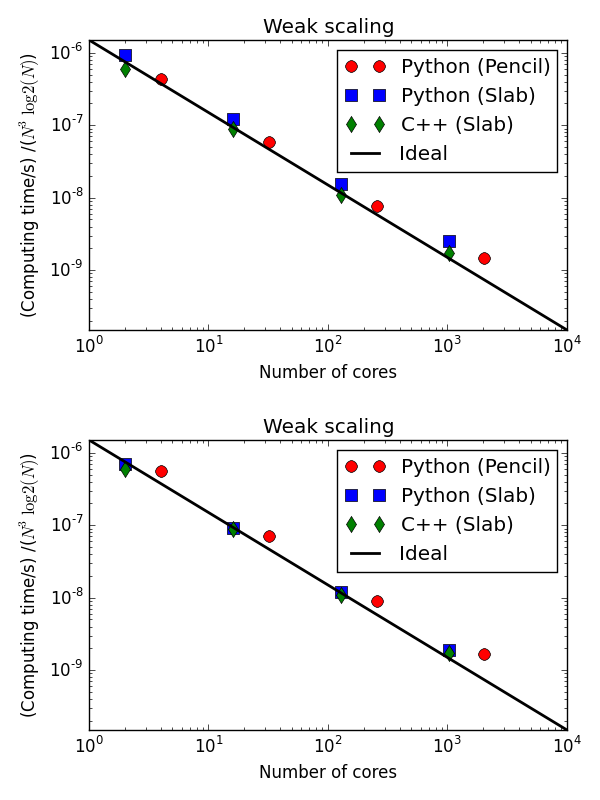}}
\caption{Weak scaling of various versions of the DNS solver. The slab decomposition uses $4 \cdot 64^3$ nodes per core, whereas the pencil decomposition uses $2 \cdot 64^3$. The C++ solver uses slab decomposition and MPI communication is performed through the FFTW library. The top figure is for a standard scientific Python solver, whereas the lower figure has some key routines optimized by Cython.  \DUrole{label}{weak}}
\end{figure}\begin{figure}[bht]\noindent\makebox[\columnwidth][c]{\includegraphics[scale=0.50]{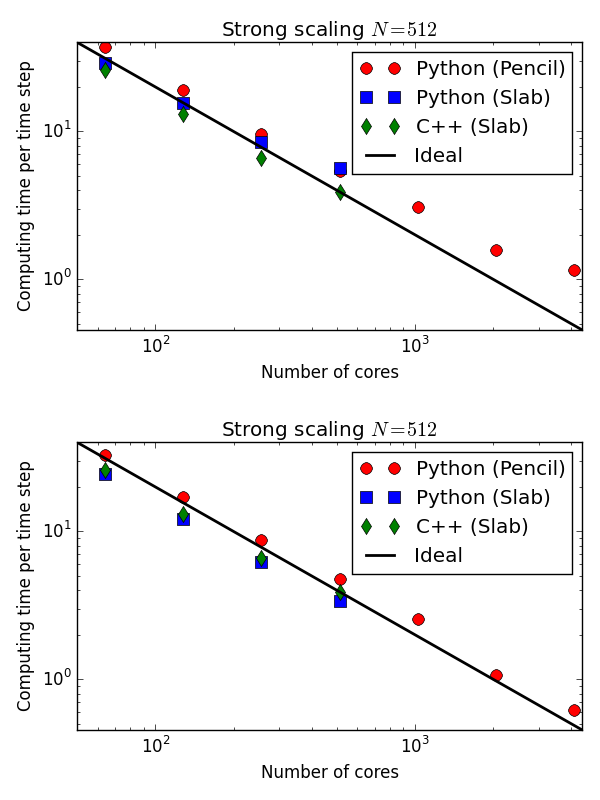}}
\caption{Strong scaling of various versions of the DNS solver. The C++ solver uses slab decomposition and MPI communication is performed through the FFTW library. The top figure is for a standard scientific Python solver, whereas the lower figure has some key routines optimized by Cython. \DUrole{label}{strong}}
\end{figure}

Next the weak scaling of the solver is tested by running the case for increasing number of processors, keeping the number of mesh nodes per CPU constant. Since the FFT is known to scale with problem size as $N \log_2 N$, and  assuming further that FFT is the major cost, the ideal weak scaling computing time should then scale proportional to $\log_2 N$. The upper panel of Figure \DUrole{ref}{weak}, shows the scaling of the scientific Python solver, both with \emph{slab} and \emph{pencil} decomposition, compared also with the C++ solver. The \emph{slab} solver uses mesh sizes of $N=(2, 16, 128, 1024)$, whereas the \emph{pencil} solver uses mesh sizes of $N=(4, 32, 256, 2048)$. The scientific Python solver is evidently 30-40 \% slower, but scaling is good - indicating that the MPI communications are performing at the level of C++. The lower panel of Figure \DUrole{ref}{weak} shows the performance of the solver when certain routines, most notably the cross product and the for-loop in the routines \emph{fftn\_mpi/ifftn\_mpi}, have been computed with Cython. The results show that the Python solver now operates very close to the speed of pure C++, and the scaling is equally good. Note that the largest simulations in Figure \DUrole{ref}{weak} are using a computational box of size $2048^3$ - approximately 8 billion mesh nodes.

Strong scaling is tested for a computational box of size $512^3$, for a various number of processors larger than 64. For \emph{slab} decomposition the maximum number of CPUs is now 512, whereas for \emph{pencil} $512^2$ CPUs can be used. The top panel of Figure \DUrole{ref}{strong} shows the performance of the scientific Python solvers. Evidently, the performance is degrading when the number of mesh nodes per CPU becomes lower and the number of processors increases. The main reason for this poor performance can be found in the implementation of the 3D FFT, where there is a for-loop over the number of processors. When this for-loop (as well as a few other routines) is moved to Cython, we observe very good strong scaling, even better than the C++ implementation that is using MPI directly from within FFTW.

To further elaborate on the performance of the code, we note that the open source pseudo-spectral C++ solver \cite{Tarang} has been benchmarked on exactly the same computer (Shaheen). Furthermore, Tarang is using the same dealiasing technique and the same 4th order Runge-Kutta integrator as we are, which should open up for direct comparison of computational efficiency. In Figure 2 of \cite{Tarang} it is shown that a computational box of size $1024^3$ is running with 512 CPUs at approximately 50 seconds per time step. In the lower panel of Figure \DUrole{ref}{weak}, we see that the current optimized Cython solver is running the same box ($1024^3$) with twice as many CPUs (1024) at approximately 20 seconds per time step. Assuming perfect strong scaling (which may be unfair considering Figure 2 of \cite{Tarang}) this would correspond to 40 seconds per time step using half the number of CPUs, which is actually 20 \% faster than Tarang.

\section{Conclusions%
  \label{conclusions}%
}

In this paper we show that it is possible to write a very good solver for direct numerical simulations of turbulent flows directly in Python, with nothing more than standard modules like NumPy, SciPy and MPI for Python (mpi4py). We also show that it is possible to get a fully competitive solver, that runs with the speed of C on thousands of processors with billions of unknowns, but then it is necessary to move a few computationally heavy routines from NumPy's ufuncs to Cython or Numba. The current paper discusses only the triply periodic domain, suitable for studying isotropic turbulence. However, the use of Python/Cython for studying turbulence is not limited to only this configuration and work is currently in progress to develop efficient Python/Cython solvers for flows with one or two inhomogeneous directions.

\section{Acknowledgements%
  \label{acknowledgements}%
}

This work is supported by the 4DSpace Strategic Research Initiative at the University of Oslo, and a Center of Excellence grant from the Research Council of Norway to the Center for Biomedical Computing at Simula Research Laboratory.








\end{document}

%% file: page_numbers.tex
\setcounter{page}{31}